# Crystal facet orientation and temperature dependence of charge and spin Hall effects in noncollinear antiferromagnet: A first-principles investigation

Meng Zhu[1], Xinlu Li[1*], Fanxing Zheng[1], Jianting Dong[1], Ye Zhou[1], Kun Wu[1] and Jia Zhang[1*]

[1]School of Physics and Wuhan National High Magnetic Field Center,
Huazhong University of Science and Technology, 430074 Wuhan, China.

* lixinlu@hust.edu.cn

* jiazhang@hust.edu.cn

## Abstract

Noncollinear antiferromagnets (nc-AFMs) have attracted increasing research attention in spintronics due to their unique spin structures and fascinating charge and spin transport properties. By using first-principles calculations, we comprehensively investigate the charge and spin Hall effects in representative noncollinear antiferromagnet $Mn_3Pt$. Our study reveals that the Hall effects in nc-AFMs are critically dependent on the crystal facet orientation and temperature. For (001) orientated $Mn_3Pt$, each charge and spin Hall conductivity element is comprised of both time reversal odd (*T*-odd) and even (*T*-even) contribution, associated with longitudinal conductivity, which leads to sizable and highly anisotropic Hall conductivity. The temperature dependence of charge and spin Hall conductivity has been elucidated by considering both phonon and spin disorder scattering. The scaling relations between Hall conductivity and longitudinal conductivity have also been investigated. The existence of prominent spin Hall effect in nc-AFMs may generate spin current with $S_z$ spin polarization, which is advantageous for field free switching of perpendicular magnetization. Our work may provide unambiguous understanding on the charge and spin transport in noncollinear antiferromagnets and pave their way for applications in antiferromagnetic spintronics.

## Introduction

Noncollinear antiferromagnets (nc-AMFs), such as $Mn_3X$ (X = Ga, Ge, Sn, Pt, Sb, etc.,) [1–4] and antiperovskite $Mn_3XN$ (X = Ni, Ga, Sn, etc.,) [5–8], etc., have attracted intensive research attention in the field of spintronics. The fascinating transport properties, including anomalous Hall effect (AHE) [9–11], spin-polarized transport [12–14], and tunneling magnetoresistance effect [15–17], make them promising for next-generation spintronic devices, such as magnetic random access memory (MRAM) [18–20].

In recent years, significant advancements have been made in the study of charge and spin Hall effects in nc-AFMs. Notably, the AHE has been theoretically predicted and experimentally observed [9,21–23] in nc-AFMs with vanishing small magnetization, challenging the conventional view that AHE is directly proportional to magnetization in traditional ferromagnets. Furthermore, spin Hall effects have also been theoretically predicted [12,24] even in the absence of spin orbit coupling and recently observed in $Mn_3Sn$ [13]. Up to now, most of the existing theoretical works on nc-AFMs investigate the individual charge and spin Hall conductivity components [22,24]. In addition, despite the temperature effect has been thoroughly studied in ferromagnets [25,26] and collinear antiferromagnets [27,28], it still remains unclear how temperature may impact the charge and spin transport properties of nc-AFMs, which makes the comparison between experimental and theoretical results unfeasible.

In this work, we leverage linear response symmetry analysis and tensor transformation relations to gain insights into the fundamental connections between magnetic symmetry, charge and spin transport in the typical nc-AFM $Mn_3Pt$. Furthermore, we use first principles calculations to shed light on the finite temperature effect on the charge and spin transport properties in $Mn_3Pt$.

## Computational Method

The electronic structure of cubic Mn₃Pt with experimental lattice constant of $a= 3.833$ Å [29] has been calculated self-consistently on the basis of local spin-density approximation (LSDA) as parameterized by Vosko et al. [30]. A wave-function expansion with angular momentum cutoff $l_{max} = 3$ is used. To investigate the transport properties, we consider two main temperature dependent scattering mechanisms, i.e., lattice vibrations (phonon) and spin disorder (spin fluctuation) scattering by employing the coherent phase approximation (CPA) alloy analogy method implemented in SPR-KKR code [26,31–33], assuming a frozen potential for the atoms [34]. Here the lattice vibrations are treated by using 14 displacement vectors with the length set to reproduce the temperature dependent root-mean-square displacement given by the Debye's theory [35], where the Debye temperature $\Theta_D = 357K$ estimated from the weighted average of the consisted elements is used. For the spin disorder scattering, we adopt temperature-dependent magnetization $M(T)$ taken from experimental data [36], where the spins are allowed to fluctuate with equal probability on a regular grid of 60 polar angles $\theta$ and 5 azimuthal angles $\varphi$. The charge and spin Hall conductivity tensors of Mn₃Pt are calculated within the linear-response theory based on the Kubo-Středa formula implemented in the Korringa-Kohn-Rostoker Green function (KKR-GF) method [37–39].

**Results and discussions**

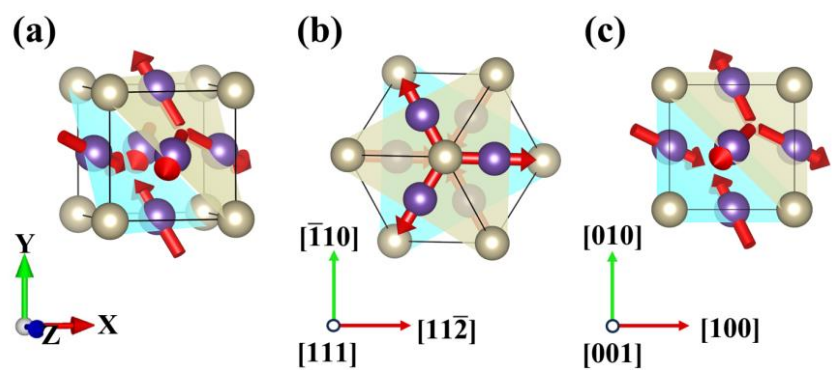

FIG. 1. (a) crystal structure and spin configuration of cubic nc-AFM Mn₃Pt. (b) and (c) are the top views of Mn₃Pt from (111) and (001) planes. The red, green, and blue arrows along different crystallographic axes indicate the $x$, $y$ and $z$ axes in the Cartesian coordinate. (b) and (c) are denoted as config1 and config2, which can be used to study the charge and spin Hall effect for (111) and (001) orientated Mn$_3$Pt, respectively.

**Symmetry-imposed charge and spin Hall conductivity tensor**

The crystal and spin configurations of the typical nc-AFM Mn$_3$Pt with space group of $Pm\bar{3}m$ (No. 221) and magnetic space group of $R\bar{3}m$ (No. 166.101) are depicted in Fig. 1a. We first analyze the symmetry-imposed charge and spin Hall conductivity tensors [40,41]. In linear response theory, the charge and spin current density propagates along $i$ direction by applying an electric field $\boldsymbol{E}$ can be written as:

$$J_i = \sum_j \sigma_{ij} E_j \qquad (i, j \in \{x, y, z\}) \tag{1a}$$

$$J_i^k = \sum_j \sigma_{ij}^k E_j \qquad (i,j,k \in \{x, y, z\}) \tag{1b}$$

where the charge Hall conductivity (CHC) $\sigma_{ij}$ is a tensor of rank two, and the spin Hall conductivity $\sigma_{ij}^k$ is a third-rank tensor with $k$ index indicating the spin-polarization of the spin current.

The most compact linear response charge and spin Hall tensors of Mn$_3$Pt can be obtained in Cartesian coordinate by setting $z$ axis along [111], $x$, $y$ along [11$\bar{2}$] and [$\bar{1}$10] crystallographic axes denoted as config1 in Fig. 1b. This configuration can be adopted for investigating the charge and spin response for (111) oriented Mn$_3$Pt film by applying in-plane electric field. The symmetry-imposed charge and spin hall conductivity tensors for Mn$_3$Pt (111) in config1 have been listed in Table 1. Hereafter we use $\tilde{\sigma}$ and $\bar{\sigma}$ to denote the $T$-odd (time reversal odd) and $T$-even (time reversal even) conductivity tensors. Specifically, the $T$-odd indicates a sign change of conductivity upon time reversal, while $T$-even denotes an invariant conductivity by time reversal operation.

As shown in Table 1, for Mn$_3$Pt(111) in config1, the charge Hall conductivity contains only nonzero $T$-odd term $\tilde{\sigma}_{xy}$. This charge hall conductivity in nc-AFMs shares the same origin as intrinsic anomalous Hall conductivity in conventional ferromagnetic metals. The $T$-odd charge Hall conductivity we calculate for Mn$_3$Pt(111) in config1 at 0 K with a constant band energy broadening of $10^{-4}$ Ry is 94 $(\Omega \cdot cm)^{-1}$, which is comparable to previous theoretical value based on Berry phase method (98 $\Omega^{-1}cm^{-1}$) [22]. For spin Hall effect, each spin conductivity element for Mn$_3$Pt in config1 contains either $T$-odd or $T$-even term. The $T$-odd spin Hall conductivity, also known as "magnetic spin Hall", is only present in magnetic materials, while the $T$-even spin Hall conductivity shares the same origin as nonmagnetic heavy metal like Pt, Ta, *etc*, [42]. Using the same band energy broadening, the calculated $T$-even spin Hall conductivity $\bar{\sigma}_{xy}^{x}$ for Mn$_3$Pt(111) in config1 at 0 K is 50 $\frac{\hbar}{e}\Omega^{-1}cm^{-1}$, in good agreement with the previous result obtained by Berry phase method (66 $\frac{\hbar}{e}\Omega^{-1}cm^{-1}$) [22]. It's worth noting that, for Mn$_3$Pt(111) in config1 the symmetry-imposed spin Hall conductivity elements $\sigma_{zx}^{x,y}$, $\sigma_{zy}^{x,y}$ are nonzero but $\sigma_{zx}^{z}$ and $\sigma_{zy}^{z}$ are zero, which suggests that the spin current propagates perpendicular to Mn$_3$Pt(111) plane by applying in-plane electric field could contain $S_x$ and $S_y$ spin-polarization components, but no out-of-plane $S_z$ component.

The charge and spin Hall conductivity tensors for other crystal facet orientation can be obtained by performing matrix transforming. For instance, by converting from the conductivity tensors for Mn$_3$Pt(111) in config1 to Mn$_3$Pt(001) in coordinate system config2 by setting [001] as $z$ axis, [100] as $x$ and [010] as $y$ axis as depicted in Fig. 1c, the transformation matrix $M$ between those two configurations is as follows:

$$M = \begin{pmatrix} \frac{\sqrt{6}}{6} & -\frac{\sqrt{2}}{2} & \frac{\sqrt{3}}{3} \\ \frac{\sqrt{6}}{6} & \frac{\sqrt{2}}{2} & \frac{\sqrt{3}}{3} \\ -\frac{\sqrt{6}}{3} & 0 & \frac{\sqrt{3}}{3} \end{pmatrix} \quad (2)$$

The charge conductivity tensors $\sigma_2$ for config2 can then be calculated based on $\sigma_1$ for config1 by performing the following matrix transformation [23]:

$$\sigma_2 = M\sigma_1 M^{-1} \quad (3)$$

The symmetry-imposed $T$-odd and $T$-even charge conductivity tensors for config2 can be expressed in terms of charge conductivity for config1 as:

$$\sigma_2^{(odd)} = \begin{pmatrix} 0 & -\frac{\sqrt{3}}{3}\tilde{\sigma}_{xy} & \frac{\sqrt{3}}{3}\tilde{\sigma}_{xy} \\ \frac{\sqrt{3}}{3}\tilde{\sigma}_{xy} & 0 & -\frac{\sqrt{3}}{3}\tilde{\sigma}_{xy} \\ -\frac{\sqrt{3}}{3}\tilde{\sigma}_{xy} & \frac{\sqrt{3}}{3}\tilde{\sigma}_{xy} & 0 \end{pmatrix} \quad \sigma_2^{(even)} = \begin{pmatrix} \frac{2\bar{\sigma}_{xx}+\bar{\sigma}_{zz}}{3} & \frac{\bar{\sigma}_{zz}-\bar{\sigma}_{xx}}{3} & \frac{\bar{\sigma}_{zz}-\bar{\sigma}_{xx}}{3} \\ \frac{\bar{\sigma}_{zz}-\bar{\sigma}_{xx}}{3} & \frac{2\bar{\sigma}_{xx}+\bar{\sigma}_{zz}}{3} & \frac{\bar{\sigma}_{zz}-\bar{\sigma}_{xx}}{3} \\ \frac{\bar{\sigma}_{zz}-\bar{\sigma}_{xx}}{3} & \frac{\bar{\sigma}_{zz}-\bar{\sigma}_{xx}}{3} & \frac{2\bar{\sigma}_{xx}+\bar{\sigma}_{zz}}{3} \end{pmatrix} \quad (4)$$

where on the right hand side of the above eq.(4), $\tilde{\sigma}_{xy}$ is the $T$-odd charge Hall conductivity, $\bar{\sigma}_{xx}$ and $\bar{\sigma}_{zz}$ are the anisotropic longitudinal conductivity for Mn$_3$Pt(111) in config1. It can be seen that, in contrast to the sole $T$-odd Hall conductivity for Mn$_3$Pt(111), each charge Hall conductivity element for Mn$_3$Pt(001) is comprised of both $T$-odd and $T$-even contributions. More importantly, the $T$-even charge Hall conductivity for Mn$_3$Pt (001) is associated with the anisotropic longitudinal conductivity. Therefore, it will result in a relatively large and highly anisotropic charge Hall conductivity at finite temperature for Mn$_3$Pt (001) as we will discuss later.

Similarly, the spin Hall conductivity elements for Mn$_3$Pt (001) in config2 $\sigma_{2,ij}^{k}$ can be obtained from the spin Hall conductivities for Mn$_3$Pt (111) in config1 $\sigma_{1,lm}^{n}$ by the following tensor transformation [23]:

$$\sigma_{2,ij}^{k} = \sum_{l,m,n} M_{il} M_{jm} M_{kn} \sigma_{1,lm}^{n} \quad (5)$$

For instance, the $T$-odd and $T$-even spin Hall conductivity elements $\sigma_{zx}^{x,y,z}$ for config2 can be expressed in terms of the spin Hall conductivity for config1 as follows:

$$\sigma_{2,zx}^{x(odd)} = \frac{1}{9}(\sqrt{6}\tilde{\sigma}_{xx}^{x} - \sqrt{3}\tilde{\sigma}_{xx}^{z} - \sqrt{3}\tilde{\sigma}_{xz}^{x} + 2\sqrt{3}\tilde{\sigma}_{zx}^{x} + \sqrt{3}\tilde{\sigma}_{zz}^{z})$$
$$\sigma_{2,zx}^{y(odd)} = \frac{1}{9}(-2\sqrt{6}\tilde{\sigma}_{xx}^{x} - \sqrt{3}\tilde{\sigma}_{xx}^{z} - \sqrt{3}\tilde{\sigma}_{xz}^{x} - \sqrt{3}\tilde{\sigma}_{zx}^{x} + \sqrt{3}\tilde{\sigma}_{zz}^{z}) \quad (6a)$$
$$\sigma_{2,zx}^{z(odd)} = \frac{1}{9}(\sqrt{6}\tilde{\sigma}_{xx}^{x} - \sqrt{3}\tilde{\sigma}_{xx}^{z} + 2\sqrt{3}\tilde{\sigma}_{xz}^{x} - \sqrt{3}\tilde{\sigma}_{zx}^{x} + \sqrt{3}\tilde{\sigma}_{zz}^{z})$$

$$\sigma_{2,zx}^{x(even)} = \frac{1}{3}(\sqrt{2}\bar{\sigma}_{xx}^{y} + \bar{\sigma}_{xz}^{y} - \bar{\sigma}_{yx}^{z})$$
$$\sigma_{2,zx}^{y(even)} = \frac{1}{3}(-\bar{\sigma}_{xz}^{y} - \bar{\sigma}_{yx}^{z} + \bar{\sigma}_{zx}^{y}) \quad (6b)$$
$$\sigma_{2,zx}^{z(even)} = \frac{1}{3}(-\sqrt{2}\bar{\sigma}_{xx}^{y} - \bar{\sigma}_{yx}^{z} - \bar{\sigma}_{zx}^{y})$$

where the quantities on the right hand side of the above eqs. (6a) and (6b) are the corresponding *T*-odd and *T*-even spin conductivity elements for config 1. It can be observed from the above equations as well as the symmetry-imposed spin conductivity tensors listed in Table I that, all the spin Hall conductivity elements for Mn$_3$Pt (001) are comprised of both *T*-odd and *T*-even terms. What's more important, $\sigma_{zx}^{x,y,z}$ and $\sigma_{zy}^{x,y,z}$ are all nonzero which indicates that Mn$_3$Pt(001) may serve as an efficient spin current source with three spin-polarization ($S_x$, $S_y$ and $S_z$) useful for field-free switching of perpendicular magnetization through spin-orbit torque mechanism. Since Mn$_3$Pt(111) is relatively trivial (it does not contain *T*-even charge Hall contribution and there is no spin Hall conductivity for generating $S_z$ spin current), in the following section we will concentrate on the charge and spin transport in Mn$_3$Pt(001).

TABLE I. The symmetry-imposed charge and spin conductivity tensors for Mn$_3$Pt. $\tilde{\sigma}$ and $\bar{\sigma}$ are used to denote the *T*-odd and *T*-even conductivity tensors. The charge and spin conductivity in the last column are our calculation results at 0 K with an imaginary energy $10^{-4}$ Ry. The units for charge Hall conductivity and spin Hall conductivity are in $(\Omega \cdot cm)^{-1}$ and $(\frac{\hbar}{2e})(\Omega \cdot cm)^{-1}$, respectively.

| | | | *T*-Odd | *T*-Even | Calculation result |
|---|---|---|---|---|---|
| | Charge conductivity | $\sigma$ | $\begin{pmatrix} 0 & \tilde{\sigma}_{xy} & 0 \\ -\tilde{\sigma}_{xy} & 0 & 0 \\ 0 & 0 & 0 \end{pmatrix}$ | $\begin{pmatrix} \bar{\sigma}_{xx} & 0 & 0 \\ 0 & \bar{\sigma}_{xx} & 0 \\ 0 & 0 & \bar{\sigma}_{zz} \end{pmatrix}$ | $\begin{pmatrix} 481061 & 94 & 0 \\ -94 & 481061 & 0 \\ 0 & 0 & 482919 \end{pmatrix}$ |
| Config1 /Mn$_3$Pt (111) | Spin conductivity | $\sigma^x$ | $\begin{pmatrix} \tilde{\sigma}_{xx}^x & 0 & \tilde{\sigma}_{xz}^x \\ 0 & -\tilde{\sigma}_{xx}^x & 0 \\ \tilde{\sigma}_{zx}^x & 0 & 0 \end{pmatrix}$ | $\begin{pmatrix} 0 & \bar{\sigma}_{xy}^x & 0 \\ \bar{\sigma}_{xy}^x & 0 & \bar{\sigma}_{yz}^x \\ 0 & \bar{\sigma}_{zy}^x & 0 \end{pmatrix}$ | $\begin{pmatrix} -43096 & 100 & 72062 \\ 100 & 43096 & -132 \\ 63390 & -176 & 0 \end{pmatrix}$ |
| | | $\sigma^y$ | $\begin{pmatrix} 0 & -\tilde{\sigma}_{xx}^x & 0 \\ -\tilde{\sigma}_{xx}^x & 0 & \tilde{\sigma}_{xz}^x \\ 0 & \tilde{\sigma}_{zx}^x & 0 \end{pmatrix}$ | $\begin{pmatrix} \bar{\sigma}_{xy}^x & 0 & -\bar{\sigma}_{yz}^x \\ 0 & -\bar{\sigma}_{xy}^x & 0 \\ -\bar{\sigma}_{zy}^x & 0 & 0 \end{pmatrix}$ | $\begin{pmatrix} 100 & 43096 & 132 \\ 43096 & -100 & 72062 \\ 176 & 63390 & 0 \end{pmatrix}$ |
| | | $\sigma^z$ | $\begin{pmatrix} \tilde{\sigma}_{xx}^z & 0 & 0 \\ 0 & \tilde{\sigma}_{xx}^z & 0 \\ 0 & 0 & \tilde{\sigma}_{zz}^z \end{pmatrix}$ | $\begin{pmatrix} 0 & \bar{\sigma}_{xy}^z & 0 \\ -\bar{\sigma}_{xy}^z & 0 & 0 \\ 0 & 0 & 0 \end{pmatrix}$ | $\begin{pmatrix} 4602 & 7 & 0 \\ -7 & 4602 & 0 \\ 0 & 0 & 4497 \end{pmatrix}$ |
| Config2 /Mn₃Pt (001) | Charge conductivity | $\sigma$ | $\begin{pmatrix} 0 & \tilde{\sigma}_{xy} & -\tilde{\sigma}_{xy} \\ -\tilde{\sigma}_{xy} & 0 & \tilde{\sigma}_{xy} \\ \tilde{\sigma}_{xy} & -\tilde{\sigma}_{xy} & 0 \end{pmatrix}$ | $\begin{pmatrix} \bar{\sigma}_{xx} & \bar{\sigma}_{xy} & \bar{\sigma}_{xy} \\ \bar{\sigma}_{xy} & \bar{\sigma}_{xx} & \bar{\sigma}_{xy} \\ \bar{\sigma}_{xy} & \bar{\sigma}_{xy} & \bar{\sigma}_{xx} \end{pmatrix}$ | $\begin{pmatrix} 481681 & 674 & 565 \\ 565 & 481681 & 674 \\ 674 & 565 & 481681 \end{pmatrix}$ |
| | Spin conductivity | $\sigma^x$ | $\begin{pmatrix} \tilde{\sigma}^x_{xx} & \tilde{\sigma}^x_{xy} & \tilde{\sigma}^x_{xy} \\ \tilde{\sigma}^x_{yx} & \tilde{\sigma}^x_{yy} & \tilde{\sigma}^x_{yz} \\ \tilde{\sigma}^x_{yx} & \tilde{\sigma}^x_{yz} & \tilde{\sigma}^x_{yy} \end{pmatrix}$ | $\begin{pmatrix} 0 & \bar{\sigma}^x_{xy} & -\bar{\sigma}^x_{xy} \\ \bar{\sigma}^x_{yx} & \bar{\sigma}^x_{yy} & \bar{\sigma}^x_{yz} \\ -\bar{\sigma}^x_{yx} & -\bar{\sigma}^x_{yz} & -\bar{\sigma}^x_{yy} \end{pmatrix}$ | $\begin{pmatrix} 78231 & 3684 & 3891 \\ -1312 & -35216 & -2613 \\ -1125 & -2647 & -35105 \end{pmatrix}$ |
| | | $\sigma^y$ | $\begin{pmatrix} \tilde{\sigma}^x_{yy} & \tilde{\sigma}^x_{yx} & \tilde{\sigma}^x_{yz} \\ \tilde{\sigma}^x_{xy} & \tilde{\sigma}^x_{xx} & \tilde{\sigma}^x_{xy} \\ \tilde{\sigma}^x_{yz} & \tilde{\sigma}^x_{yx} & \tilde{\sigma}^x_{yy} \end{pmatrix}$ | $\begin{pmatrix} -\bar{\sigma}^x_{yy} & -\bar{\sigma}^x_{yx} & -\bar{\sigma}^x_{yz} \\ \bar{\sigma}^x_{xy} & 0 & -\bar{\sigma}^x_{xy} \\ \bar{\sigma}^x_{yz} & \bar{\sigma}^x_{yx} & \bar{\sigma}^x_{yy} \end{pmatrix}$ | $\begin{pmatrix} -35105 & -1125 & -2647 \\ 3891 & 78231 & 3684 \\ -2613 & -1312 & -35216 \end{pmatrix}$ |
| | | $\sigma^z$ | $\begin{pmatrix} \tilde{\sigma}^x_{yy} & \tilde{\sigma}^x_{yz} & \tilde{\sigma}^x_{yx} \\ \tilde{\sigma}^x_{yz} & \tilde{\sigma}^x_{yy} & \tilde{\sigma}^x_{yx} \\ \tilde{\sigma}^x_{xy} & \tilde{\sigma}^x_{xy} & \tilde{\sigma}^x_{xx} \end{pmatrix}$ | $\begin{pmatrix} \bar{\sigma}^x_{yy} & \bar{\sigma}^x_{yz} & \bar{\sigma}^x_{yx} \\ -\bar{\sigma}^x_{yz} & -\bar{\sigma}^x_{yy} & -\bar{\sigma}^x_{yx} \\ -\bar{\sigma}^x_{xy} & \bar{\sigma}^x_{xy} & 0 \end{pmatrix}$ | $\begin{pmatrix} -35216 & -2613 & -1312 \\ -2647 & -35105 & -1125 \\ 3684 & 3891 & 78231 \end{pmatrix}$ |

## Charge Hall conductivity of Mn₃Pt: temperature effects and anisotropy

In order to evaluate the full charge or spin conductivity tensors for nc-AFMs, previous calculations usually adopted constant band energy broadening (*i.e.* constant relaxation time) approximation [12]. However, to make a direct comparison between experimental and theoretical results, finite temperature effect including phonon and spin disorder scattering should be explicitly taken into account. To better understand the transport property of nc-AFMs at finite temperature, we first calculate and plot the Bloch spectra function (BSF) for cubic Mn₃Pt along the high symmetry *k* paths in Brillouin zone. The calculated BSFs at various temperatures by considering both phonon and spin fluctuation have been shown in Fig. 2. At 0 K, the BSF of Mn₃Pt has zero energy broadening (*i.e.* infinite electron lifetime), which is identical with its band structure shown in Fig. S1 in Supplementary Material [43]. As depicted in Fig. 2(b) to Fig. 2(d), when the temperature is rising, the BSF becomes increasingly blurred. The enhancement of electron scattering with phonon and spin disorder, leads to a larger band energy broadening and a shorter electron lifetime at higher temperature can also be visible from temperature dependent density of states shown in Fig.S2 of the Supplementary Material [43].

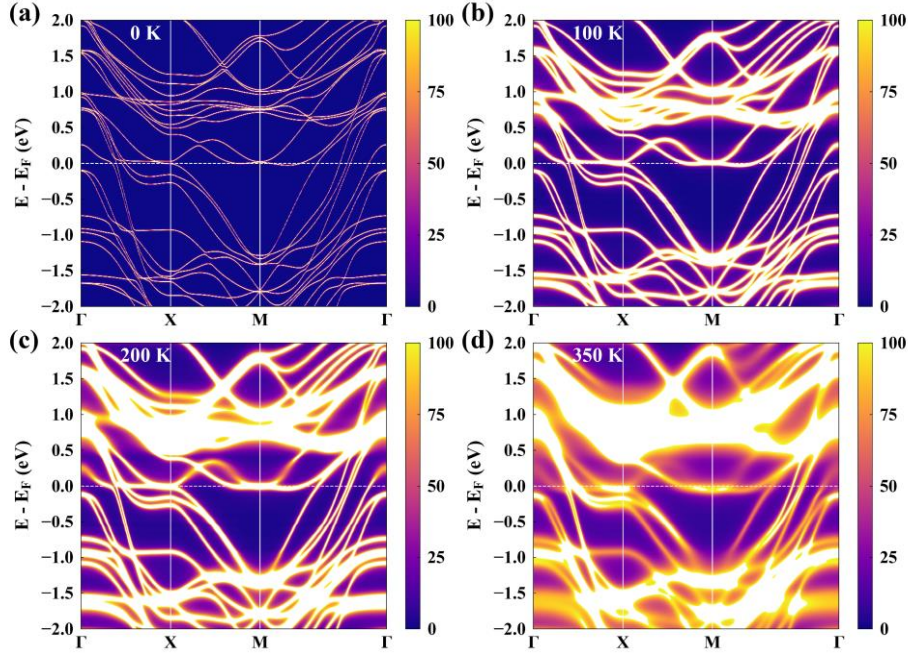

**FIG. 2.** The Bloch spectra function (BSF) of cubic Mn$_3$Pt calculated at 0 K (a), 100 K (b), 200 K (c) and 350 K (d). The horizontal white dashed lines indicate the Fermi energy.

According to the linear response theory, the *T*-odd and *T*-even charge conductivity tensors can be calculated within the constant band energy broadening approximation as follows [12,44]:

$$\sigma_{ij}^{(odd)} = e\hbar \sum_{n\neq m} [f(\varepsilon_m) - f(\varepsilon_n)] \frac{\text{Im}(\langle n|\hat{J}_i|m\rangle\langle m|\hat{v}_j|n\rangle)}{(\varepsilon_n - \varepsilon_m)^2 + (\hbar\Gamma)^2} \qquad (7a)$$

$$\sigma_{ij}^{(even)} = -e\hbar^2 \Gamma \sum_{n,m} \frac{f(\varepsilon_m) - f(\varepsilon_n)}{\varepsilon_n - \varepsilon_m} \frac{\text{Re}(\langle n|\hat{J}_i|m\rangle\langle m|\hat{v}_j|n\rangle)}{(\varepsilon_n - \varepsilon_m)^2 + (\hbar\Gamma)^2} \qquad (7b)$$

where *m* and *n* are the band indices, $f(\varepsilon)$ is the Fermi-Dirac distribution. $\hat{J}$ and $\hat{v}$ represent the electric current density and velocity operators, respectively. *e* is elementary charge, $\hbar$ is the reduced Planck constant and $\hbar\Gamma$ is the constant band energy broadening by taking into account of the finite electron lifetime due to scattering.

It's clear that, generally the *T*-odd charge Hall conductivity will decrease with the increase of temperature and band energy broadening $\hbar\Gamma$. In the clean limit $\Gamma \rightarrow 0$, the *T*-odd charge Hall conductivity $\sigma_{ij}^{(odd)}$ approaches a finite value and restores to Berry phase expression for intrinsic anomalous Hall conductivity [45]. The *T*-even charge Hall conductivity can be decomposed into intraband (*m=n*) and interband ($m \neq n$) contributions:

$$\sigma_{ij}^{(even,intra)} = \frac{e}{\Gamma} \sum_n \langle n|\hat{J}_i|n\rangle \langle n|\hat{v}_j|n\rangle \left(-\frac{\partial f(\varepsilon_n)}{\partial \varepsilon_n}\right) \quad (8a)$$

$$\sigma_{ij}^{(even,inter)} = -e\hbar^2 \Gamma \sum_{n \neq m} \frac{f(\varepsilon_m) - f(\varepsilon_n)}{\varepsilon_n - \varepsilon_m} \frac{\text{Re}(\langle n|\hat{J}_i|m\rangle \langle m|\hat{v}_j|n\rangle)}{(\varepsilon_n - \varepsilon_m)^2 + (\hbar\Gamma)^2} \quad (8b)$$

The *T*-even charge conductivity including longitudinal conductivity ($\sigma_{ii}$) also decreases with the increase of temperature. In the clean limit $\Gamma \to 0$, the interband term $\sigma_{ij}^{(even,inter)}$ gradually vanishes, while the intraband contribution $\sigma_{ij}^{(even,intra)}$ becomes dominating and diverges as a function of $1/\Gamma$.

We then calculate the temperature dependent charge conductivity by including phonon and spin disorder scatterings by using alloy analogy model as implemented in Green's function based SPR-KKR method [32]. We first compare the longitudinal resistivity $\rho_{xx}$ with the available experimental results for bulk single crystal Mn$_3$Pt [46]. As shown in Fig. 3a, the calculated electrical resistivity by considering both lattice vibration and spin fluctuation agrees well with the experimental result. When only the lattice vibration effect is included, the phonon resistivity $\rho_{xx}^{vib}$ scales almost linearly with temperature *T*. In the low temperature regime ($T < 200\,\text{K}$), the spin fluctuation resistivity $\rho_{xx}^{flu}$ is comparable to $\rho_{xx}^{vib}$, and $\rho_{xx}^{flu}$ is dominant over $\rho_{xx}^{vib}$ in the high temperature regime ($T > 220\,\text{K}$), which indicates that the spin disorder scattering in nc-AFMs is vital for charge and spin transport at finite temperatures.

Experimentally, the anomalous Hall resistivity of nc-AFMs has usually been measured and evaluated by $\frac{1}{2}[\rho_{xy}^+(H=0) - \rho_{xy}^-(H=0)]$, which only contains the *T*-odd term, where $\rho_{xy}^{\pm}(H=0)$ refer to Hall resistivity measured at zero magnetic field after applying opposite magnetic field [10,46]. As shown in Fig. 3b, the calculated *T*-odd charge Hall conductivity for Mn$_3$Pt (001) agrees well with the experimental result in a wide temperature window. At room temperature (T=300K), the calculated *T*-odd hall conductivity is 33 $(\Omega \cdot cm)^{-1}$ which is close to the experimental value of 25 $(\Omega \cdot cm)^{-1}$ [46].

As shown in Fig. 3b, both calculated *T*-odd and *T*-even charge Hall conductivity decrease with increasing temperature. At the same temperature, the *T*-even charge Hall conductivity $\bar{\sigma}_{xy}$, which is associated with the anisotropic longitudinal transport, is dozens of times larger than the *T*-odd charge Hall conductivity $\tilde{\sigma}_{xy}$. We further investigate the scaling relation between charge Hall conductivity and longitudinal conductivity $\sigma_{xx}$. As shown in Fig. 3d, the *T*-even charge Hall conductivity $\bar{\sigma}_{xy}$ can be well described by linearly scaling with $\sigma_{xx}$ as $\bar{\sigma}_{xy} \sim a\sigma_{xx} + b$, since the *T*-even conductivity shows the similar dependence on band energy broadening as indicated in eq.(7). However, the *T*-odd charge Hall conductivity $\tilde{\sigma}_{xy}$, which exhibits different dependence on band energy broadening, should be described by including both linear and quadratic scaling with $\sigma_{xx}$ as $\tilde{\sigma}_{xy} \sim a'\sigma_{xx}^2 + b'\sigma_{xx} + c'$. Such scaling relation between $\tilde{\sigma}_{xy}$ and $\sigma_{xx}$ for Mn$_3$Pt film has been reported by recent experiment [47].

The in-plane anisotropy of charge Hall conductivity for Mn$_3$Pt (001) at room temperature is further studied. As shown in Fig.3c, $\theta$ has been set to be the angle between electric field and the [100] crystal axis. By employing similar matrix transformation as shown in eq. (3), the charge Hall conductivity for Mn$_3$Pt (001) with arbitrary in-plane angle $\theta$ respecting to config2 can be obtained by the following transformation matrix:

$$M = \begin{pmatrix} \cos\theta & -\sin\theta & 0 \\ \sin\theta & \cos\theta & 0 \\ 0 & 0 & 1 \end{pmatrix} \quad (9)$$

The calculated total charge Hall conductivity $\sigma_{xy}$, the *T*-odd ($\tilde{\sigma}_{xy}$) and *T*-even ($\bar{\sigma}_{xy}$) Hall conductivity as a function of $\theta$ has been shown in Fig 3c. The anisotropic charge hall conductivity generally follows a sinusoidal function with a $\pi$ periodicity. The highly anisotropic total Hall conductivity $\sigma_{xy}$ can be attributed to the dominant $T$-even Hall conductivity $\bar{\sigma}_{xy}$. The $T$-even charge Hall conductivity reaches its maximum value 350 $\Omega^{-1}$cm$^{-1}$ at $\theta$=0, an order of magnitude larger than the $T$-odd Hall conductivity (34 $\Omega^{-1}$cm$^{-1}$).

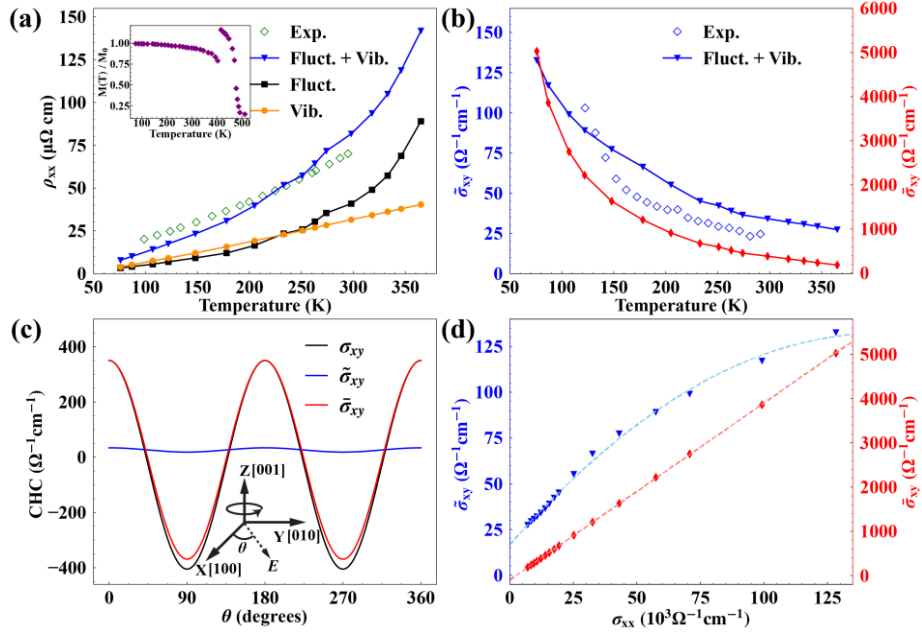

**FIG. 3.** (a) The calculated temperature dependence of longitudinal resistivity $\rho_{xx}$. The experimental longitudinal resistivity for single crystal bulk Mn$_3$Pt is shown in green open diamond for comparison [46]. The inset is the experimental $M(T)$ curve of Mn$_3$Pt we use for considering spin fluctuation scattering. (b) The calculated temperature dependent $T$-odd (blue triangles, refer to left axis) and $T$-even (red diamonds, refer to right axis) charge Hall conductivity of Mn$_3$Pt(001) in config2. $T$-odd experimental data is shown in blue open diamonds for comparison [46]. (c) The total, $T$-odd and $T$-even charge Hall conductivity as a function of $\theta$ for Mn$_3$Pt(001) at room temperature 300 K. The insect shows the definition of angle $\theta$. (d) The calculated $T$-odd (blue triangles, refer to left axis) and $T$-even (red diamonds, refer to right axis) charge Hall conductivity as a function of longitudinal conductivity $\sigma_{xx}$. The dashed lines are the fitting curves based on scaling relation $\tilde{\sigma}_{xy} \sim a'\sigma_{xx}^2 + b'\sigma_{xx} + c'$ and $\bar{\sigma}_{xy} \sim a\sigma_{xx} + b$, respectively.

# Spin Hall conductivity of Mn$_3$Pt: temperature effects and anisotropy

The spin current generated by spin Hall effect with out-of-plane $S_z$ component (spin-polarization perpendicular to film plane) is essential for field-free switching of perpendicular magnetization. However, as shown in Fig.4(a), in conventional nonmagnetic heavy metal like Pt and Ta, the spin current propagating perpendicular to film plane only have in-plane spin polarization. Recently, several experiments demonstrated that nc-AFMs can generate spin current with three spin polarization ($S_x$, $S_y$, $S_z$), which makes nc-AFMs an appealing spin current source for spin-orbit torque (SOT) applications [19,20]. The full spin conductivity tensor for Mn$_3$Pt (001) in config2 can be found in Table 1. It can be seen that all spin Hall conductivity elements, especially $\sigma_{zx}^{x,y,z}$ and $\sigma_{zy}^{x,y,z}$ are nonzero and contributed by major $T$-odd and minor $T$-even terms. This indicates that, as shown in Fig.4(b) for Mn$_3$Pt(001) film, there will be spin current with three spin polarization components ($S_x$, $S_y$, $S_z$) propagating along the $z$ direction by applying an in-plane electric field, which makes Mn$_3$Pt(001) and similar nc-AFMs a promising spin current source for SOT application.

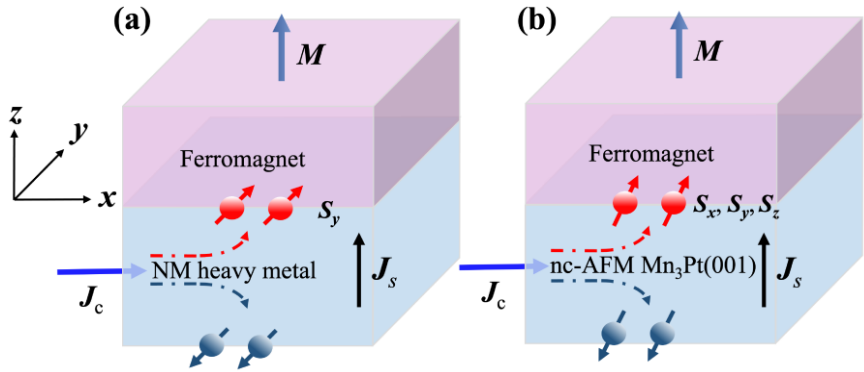

**Fig.4.** Comparison of spin Hall effect for SOT applications: (a) Conventional nonmagnetic heavy metals like Pt and Ta may generate spin current with only $S_y$ spin-polarization by applying charge current $J_c$ along $x$ direction. (b) Representative nc-AFM such as Mn$_3$Pt(001) could produce spin current with all three spin-polarization $S_x$, $S_y$, and $S_z$, which is beneficial for field-free switching of perpendicular magnetization.

Similar to charge conductivity, the linear response $T$-odd and $T$-even spin conductivity based on constant band energy broadening approximation can be obtained by replacing charge current density operator with spin current operator $\hat{J}_i^k = \frac{1}{2}\{\hat{s}_k, \hat{v}_i\}$ and interchange $T$-odd and $T$-even terms in eq.(7) as [12,44]:

$$\sigma_{ij}^{k(\text{even})} = e\hbar \sum_{n \neq m}[f(\varepsilon_m) - f(\varepsilon_n)] \frac{\text{Im}(\langle n|\hat{J}_i^k|m\rangle\langle m|\hat{v}_j|n\rangle)}{(\varepsilon_n - \varepsilon_m)^2 + (\hbar\Gamma)^2} \quad (10a)$$

$$\sigma_{ij}^{k(\text{odd})} = -e\hbar^2 \Gamma \sum_{n,m} \frac{f(\varepsilon_m) - f(\varepsilon_n)}{\varepsilon_n - \varepsilon_m} \frac{\text{Re}(\langle n|\hat{J}_i^k|m\rangle\langle m|\hat{v}_j|n\rangle)}{(\varepsilon_n - \varepsilon_m)^2 + (\hbar\Gamma)^2} \quad (10b)$$

It's clear that generally the $T$-even spin Hall conductivity $\sigma_{ij}^{k(\text{even})}$ decreases with the increase of temperature. In the clean limit $\Gamma \to 0$, the $T$-even spin Hall conductivity approaches to a finite value and can be evaluated based on Berry phase expression for intrinsic spin Hall conductivity [48]. The $T$-odd spin Hall conductivity (i.e. magnetic spin hall conductivity) $\sigma_{ij}^{k(\text{odd})}$ can be decomposed into intraband ($m=n$) and interband ($m \neq n$) contributions:

$$\sigma_{ij}^{k(\text{odd,intra})} = \frac{e}{\Gamma} \sum_n \langle n|\hat{J}_i^k|n\rangle\langle n|\hat{v}_j|n\rangle\left(-\frac{\partial f(\varepsilon_n)}{\partial \varepsilon_n}\right) \quad (11a)$$

$$\sigma_{ij}^{k(\text{odd,inter})} = -e\hbar^2 \Gamma \sum_{n \neq m} \frac{f(\varepsilon_m) - f(\varepsilon_n)}{\varepsilon_n - \varepsilon_m} \frac{\text{Re}(\langle n|\hat{J}_i^k|m\rangle\langle m|\hat{v}_j|n\rangle)}{(\varepsilon_n - \varepsilon_m)^2 + (\hbar\Gamma)^2} \quad (11b)$$

The $T$-odd spin Hall conductivity also decreases when the temperature increases. In the clean limit $\Gamma \to 0$, the interband term $\sigma_{ij}^{k(\text{odd,inter})}$ gradually reduces to zero, while the intraband contribution $\sigma_{ij}^{k(\text{odd,intra})}$ becomes dominating and diverges as a function of $1/\Gamma$.

We then calculate the temperature dependent spin Hall conductivity (SHC) by including both phonon and spin disorder scatterings. It can be seen from Table I that the spin Hall conductivity elements for Mn$_3$Pt(001) in config2 are mostly contributed by the $T$-odd term with a small portion of $T$-even term. Fig. 5(a) to Fig. 5(c) present the temperature dependency of three typical spin Hall conductivity elements $\sigma_{zx}^x$, $\sigma_{zx}^y$ and $\sigma_{zx}^z$ for Mn$_3$Pt(001). At low temperature, the SHCs are all at the order of $10^3 \hbar/2e(\Omega \cdot cm)^{-1}$. With the increase of temperature, the absolute values of all the three SHCs drastically decrease and drop to around $10^2 \hbar/2e(\Omega \cdot cm)^{-1}$ at room temperature 300 K.

The scaling relation between spin Hall conductivity with longitudinal conductivity $\sigma_{xx}$ has also been investigated. As it is shown in Fig. 5(d-g), all three spin Hall conductivity elements $\sigma_{zx}^{x,y,z}$ follow linear scaling with $\sigma_{xx}$ and can be well described as $\sigma_{zx}^k \sim a\sigma_{xx} + b$. Such linear scaling behavior of spin Hall conductivity can be understood based on the fact that both the dominating T-odd spin Hall conductivity and the T-even longitudinal conductivity $\sigma_{xx}$ exhibit similar dependence on band energy broadening $\hbar\Gamma$ as indicated in eq.(10b) and eq.(7b).

The arbitrary in-plane angle $\theta$ dependence of $\sigma_{zx}^{x,y,z}(\theta)$ for Mn$_3$Pt(001) can be obtained from spin Hall tensor of Mn$_3$Pt(001) in config2 by performing tensor transformation by following eq.(6):

$$\sigma_{zx}^x(\theta) = \sigma_{zx}^x \cos^2\theta - \frac{\sigma_{zx}^y \sin 2\theta}{2} - \frac{\sigma_{zy}^x \sin 2\theta}{2} + \sigma_{zy}^y \sin^2\theta$$

$$\sigma_{zx}^y(\theta) = \frac{\sigma_{zx}^x \sin 2\theta}{2} + \sigma_{zx}^y \cos^2\theta - \sigma_{zy}^x \sin^2\theta - \frac{\sigma_{zy}^y \sin 2\theta}{2} \quad (12)$$

$$\sigma_{zx}^z(\theta) = \sigma_{zx}^z \cos\theta - \sigma_{zy}^z \sin\theta$$

where on the right hand side are the corresponding spin Hall conductivities for Mn$_3$Pt(001). The calculated spin Hall conductivities and corresponding spin Hall angles $\alpha^{x,y,z}$ (the spin Hall angles are defined as $\alpha^{x,y,z} = \frac{2e}{\hbar}(\frac{\sigma_{zx}^{x,y,z}}{\sigma_{xx}})$) as functions of in-plane angle $\theta$ at room temperature are shown in Fig. 5(g) to Fig. 5(i). It can be seen that $\alpha^{x,y,z}$ are all at the order of several percent. $\alpha^{x,y}(\theta)$ follow a $\pi$ periodicity, while $\alpha^z(\theta)$ has a $2\pi$ periodicity as manifested in eq.(12).

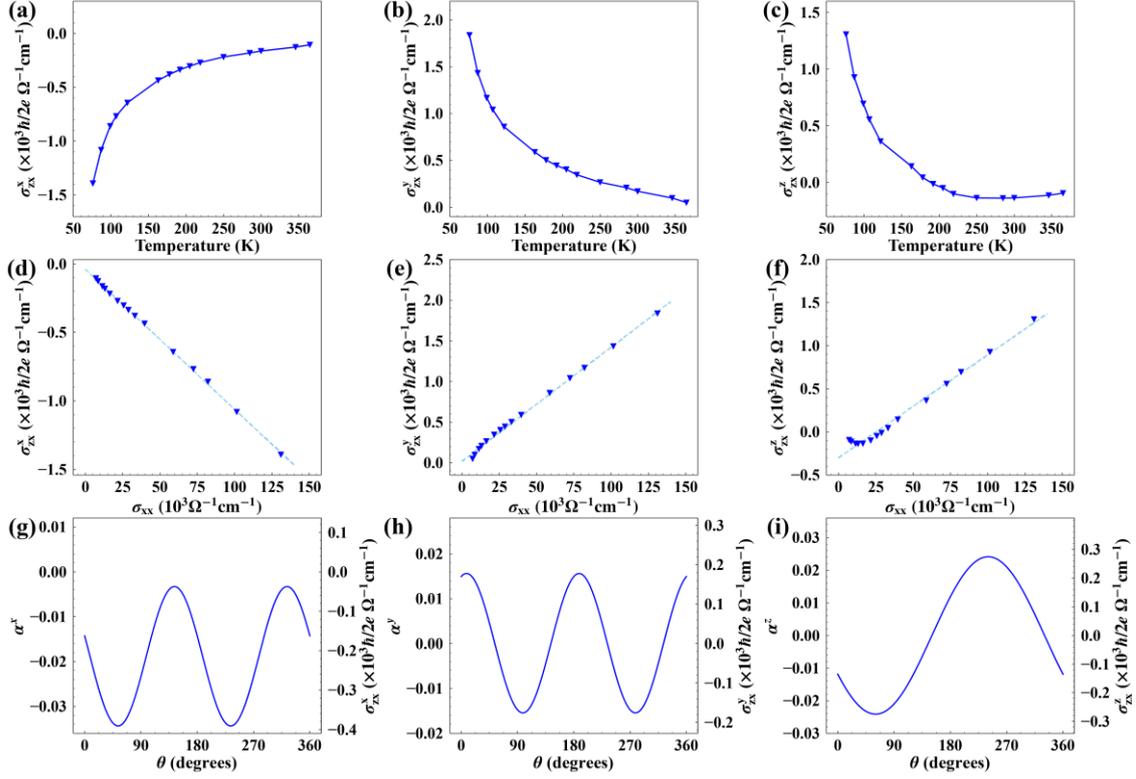

**FIG. 5.** The calculated spin Hall conductivity $\sigma_{zx}^{x}$ (a), $\sigma_{zx}^{y}$ (b) and $\sigma_{zx}^{z}$ (c) of Mn$_3$Pt(001) in config2 as a function of temperature. (d-f) are the scaling relation between $\sigma_{zx}^{x}$, $\sigma_{zx}^{y}$, $\sigma_{zx}^{z}$ and the longitudinal conductivity $\sigma_{xx}$, where the dashed lines are linear fittings by $\sigma_{zx}^{x,y,z} \sim a\sigma_{xx} + b$. (g-i) are the in-plane anisotropic spin Hall conductivity (refer to right axis) and spin Hall angle (refer to left axis) for Mn$_3$Pt(001) at room temperature (300 K).

We then focus on $\alpha^z$ and $\sigma_{zx}^{z}$ which are essential for field-free switching of perpendicular magnetization. The peak spin Hall angle $\alpha^z$ for (001) oriented Mn$_3$Pt is calculated to be 0.024, corresponding to spin Hall conductivity $\sigma_{zx}^{z}$=275 $\hbar/2e$ $\Omega^{-1}$cm$^{-1}$. As illustrated in Fig. 6, the calculated spin Hall conductivity $\sigma_{zx}^{z}$ of Mn$_3$Pt(001) for $S_z$ spin polarization is the largest among the available materials and at the same order with Mn$_3$Ir ( 143 $\hbar/2e$ $\Omega^{-1}$cm$^{-1}$ ) [49] and 2D WTe$_2$/PtTe$_2$ heterostructure ( 250 $\hbar/2e$ $\Omega^{-1}$cm$^{-1}$ ) [50]. Therefore, spin current with large $S_z$ spin polarization can be generated by spin Hall effect in Mn$_3$Pt(001), which is particularly advantageous for spin-orbit torque (SOT) applications.

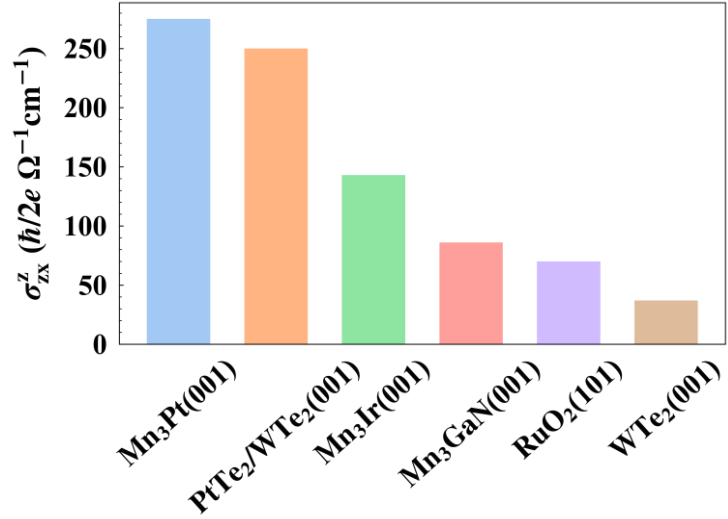

**FIG. 6.** Comparison of the calculated spin Hall conductivity $\sigma_{zx}^z$ relevant for spin current with out-of-plane $S_z$ spin polarization for Mn$_3$Pt(001) at 300 K and the reported experimental results for PtTe$_2$/WTe$_2$/ [50], Mn$_3$Ir [49], Mn$_3$GaN [51], RuO$_2$ [52] and WTe$_2$ [53].

## Summary

In summary, we have presented first principles calculations for the charge and spin Hall effects in nc-AFM Mn$_3$Pt at finite temperature by including phonon and spin disorder scatterings. The charge and spin transports are found to be critically dependent on crystal facet orientation. For Mn$_3$Pt(001) it contains both *T*-odd and *T*-even charge Hall contribution, while the *T*-even part is associated with anisotropic longitudinal transport, leading to relatively large and highly anisotropic total Hall conductivity. The *T*-even charge Hall conductivity shows linear scaling with longitudinal conductivity $\sigma_{xx}$, while the *T*-odd term should be scaled by also including quadratic term. The spin Hall conductivity tensors for Mn$_3$Pt(001) are full and each element is comprised of major *T*-odd contribution and minor *T*-even contribution, making spin Hall conductivity linearly scaling with $\sigma_{xx}$. The large spin Hall conductivity element for producing out-of-plane spin polarization $S_z$ at room temperature makes it useful for field-free switching of perpendicular magnetization in SOT applications. Our work may provide comprehensive understandings on the charge and spin Hall effect in nc-AFMs and pave the way for potential applications in spintronic devices.

## Acknowledgment

This work was supported by the National Natural Science Foundation of China (grant No. T2394475, T2394470), and the open research fund of Beijing National Laboratory for Condensed Matter Physics (No.2023BNLCMPKF010). X.L. Li was supported by the China Postdoctoral Science Foundation under Grant 2023M741269.